# Hidden integer quantum ferroelectricity in chiral Tellurium

Wei Luo[1]*, Sihan Deng[2]*, Muting Xie[2], Junyi Ji[3]†, Hongjun Xiang[2,4]† and Laurent Bellaiche[1,5]

[1]Smart Ferroic Materials Center, Physics Department and Institute for Nanoscience and Engineering, University of Arkansas Fayetteville, Arkansas 72701, USA

[2]Key Laboratory of Computational Physical Sciences (Ministry of Education), State Key Laboratory of Surface Physics, and Department of Physics, Fudan University, Shanghai 200433, China

[3]Beijing National Laboratory for Condensed Matter Physics and Institute of Physics, Chinese Academy of Sciences, Beijing 100190, China.

[4]Collaborative Innovation Center of Advanced Microstructures, Nanjing 210093, China

[5]Department of Materials Science and Engineering, Tel Aviv University, Ramat Aviv, Tel Aviv 6997801, Israel

*These authors contributed equally to this work.

†Contact author: jyji@iphy.ac.cn

†Contact author: hxiang@fudan.edu.cn

**Abstract:** Ferroelectricity is a cornerstone of functional materials research, enabling diverse technologies from non-volatile memory to optoelectronics. Recently, type-I integer quantum ferroelectricity (IQFE), unconstrained by symmetry, has been proposed and experimentally demonstrated; however, as it arises from ionic displacements of an integer lattice vector, the initial and final states are macroscopically indistinguishable, rendering the physical properties unchanged. Here, we propose for the first time the nontrivial counterpart (i.e., type-II IQFE) where the polarization difference between the initial and final states is quantized but the macroscopical properties differ. We further demonstrate the existence of type-II IQFE in bulk chiral tellurium. In few-layer tellurium, the total polarization remains nearly quantized, composed of a bulk-inherited quantum component and a small surface-induced contribution. Molecular dynamics simulations reveal surface-initiated, layer-by-layer switching driven by reduced energy barriers, explaining why ferroelectricity was observed experimentally in few-layer tellurium, but not in bulk tellurium yet. Interestingly, the chirality of the initial and final states in bulk tellurium is opposite, suggesting a novel way to control structural chirality with electric field in chiral photonics and nonvolatile ferroelectric memory devices.

**Introduction**

Ferroelectric materials have garnered considerable scientific and technological interest due to their rich array of physical properties and broad spectrum of applications. They play a pivotal role in non-volatile memory devices (FeRAM)[1-6], piezoelectric sensors and actuators[7-10], and energy harvesting technologies[11-13]. In recent years, emerging forms of ferroelectricity—such as stacking ferroelectricity[14-21], metallic ferroelectricity[22-27], integer and fractional quantum ferroelectricity[28,29]—have attracted increasing attention owing to their unconventional physical mechanisms and potential for novel functionalities. The discovery of such non-traditional ferroelectrics not only expands the known landscape of ferroelectric materials but also opens new avenues for their application beyond conventional electronic devices. However, such fascinating ferroelectric materials usually possess several different atoms.

Recent experimental studies have confirmed the presence of elemental ferroelectricity in two-dimensional (2D) bismuth[30-32]. For 2D bismuth, the ferroelectricity originates from charge transfer between different sublattices, as bismuth can exhibit both positive and negative valence states due to different local atomic environments. Very recently, the thin-film tellurium (Te) has also been experimentally identified as a single-element ferroelectric material, exhibiting high carrier mobility[33,34]. In contrast to 2D bismuth, tellurium lacks distinct sublattices and exhibits an identical local environment for each Te atom. Consequently, the ferroelectric mechanism in Te must differ from that in 2D bismuth. On the other hand, the experimentally reported Te films are relatively thick—ranging from 20 to 100 nm—raising an important question regarding the origin of the observed ferroelectricity: does it stem from a genuine bulk property, or is it a result of surface atomic reconstruction? A clear understanding of the origin of ferroelectricity is not only of fundamental scientific importance but also crucial for the development of novel ferroelectric device technologies.

Here, we present a comprehensive theoretical investigation showing that bulk chiral Te, which belongs to the nonpolar point group $D_3$, can host a hidden form of ferroelectricity—termed type-II integer quantum ferroelectricity (IQFE). We construct a toy model to illustrate how an integer quantum polarization can emerge through coupled ionic and electronic displacements in a single element system. We then demonstrate that this model is physically realized in three-dimensional (3D) chiral Te, where first-principles calculations of Wannier charge centers (WCCs) reveal an integer quantum polarization difference between the two chiral states. Using machine learning potential (MLP) based molecular dynamics (MD) simulations, we show that this polarization can be reversibly switched under an applied electric field. Extending our analysis to thin-film Te, we identify a nearly quantized polarization that arises from the inherited bulk quantum contribution combined with a surface component. The thickness dependence of this polarization, as well as the atomically resolved switching dynamics, highlight the critical role of surface atoms in enabling ferroelectric switching for thin-film Te. Our work demonstrates that the ferroelectric origin of 2D Te is inherited from its bulk counterpart and opens new avenues for exploring type-II IQFE in nonpolar symmetry groups.

**Toy model of type-II IQFE**

Before discussing the nontrivial type-II IQFE, we first introduce the trivial type-I IQFE, as illustrated in Fig. 1(a). In this case, all anions (or cations) are displaced by an integer multiple of the lattice constant, resulting in a quantized change in dipole moment between the initial and final states. Importantly, these two states are related by a pure lattice translation and are therefore structurally identical when considering the periodic boundary condition. For nontrivial type-II IQFE, the anions (or cations) are displaced by an arbitrary multiple of the lattice constant, rather than being restricted to integer or fractional values; yet, the resulting polarization remains quantized to an integer multiple of the polarization quantum. To illustrate the concept of type-II IQFE, we begin with a toy lattice model composed of a single element and governed by $D_{3h}$ point group symmetry, as depicted in Fig. 1b. Each primitive cell contains three lattice sites (denoted by red numbers 1, 2, and 3 in cell $C_1$ of Fig. 1b), and the in-plane lattice constant is denoted by $a$. We assume that each site carries a positive charge of $+ne$ and each bond center (i.e., the center of the Wannier functions, labeled as 4, 5, and 6 in Fig. 1b) carries a negative charge of $-ne$, where $n$ is an integer and $e$ is the elementary charge. Under this charge configuration, the system remains overall charge-neutral. Although the $D_{3h}$ point group is non-polar, we argue that an integer-quantized dipole moment difference of $ne*a$ along the $a$-axis may exist between the initial configuration and a final, symmetry-related configuration (atom 1 in cell $C_1$ shifts to the position of atom 1' in cell $C_2$, see Fig. 1b). This is possible because, according to the modern theory of polarization [35], the polarization of a periodic crystal is a multi-value vector; hence, the dipole moment of a non-polar crystal can be non-zero. More recently, the presence of a non-zero polarization in non-polar crystals was reconciled within the generalized Neumann's principle [36]. To clarify this point, we analyze the dipole moment along the $a$-axis by choosing site 3 in cell $C_1$ of Fig. 1b as the origin. For the initial state, the dipole moment is:

$$P_i = P_i^2 + P_i^3 + P_i^4 + P_i^1 + P_i^5 + P_i^6 = 0 + 0 + 0 + nex - ne\frac{x}{2} - ne\frac{x}{2} = 0 \quad (1)$$

Here, $P_i^1$, $P_i^2$ and $P_i^3$ represent the dipole moments contributed by sites 1, 2, and 3 (Fig. 1b), while $P_i^4$, $P_i^5$ and $P_i^6$ represent the dipole moment contributed by the WCCs. The $x$ denote the coordinate of site 1 in cell $C_1$ along the $a$-axis (see Fig. 1b). For the final state—constructed by shifting site 1 from cell $C_1$ to cell $C_2$ (1') by a displacement $\Delta d$ along the $a$-axis (as illustrated in Fig. 1b). One can see that, during the displacement of the atom 1 (in cell $C_1$), as indicated by the dashed red arrow (Fig. 1b) pointing from 1 to 1', the bonds between atom 1 and atoms 2 and 3 (within cell $C_1$) are broken. Concurrently, new bonds are formed between atom 1' and atoms 2 and 3 in the adjacent cell $C_2$, as represented by the dashed black lines in cell $C_2$. It is precisely the breaking and reformation of these bonds that lead to the shift of the Wannier centers from the left side (black balls in cell $C_1$) to the right side (dashed black circle in cell $C_2$), as shown by the dashed black arrows in Fig. 1(b). The dipole moment of the final state can be calculated as:

$$P_f = P_i^2 + P_i^3 + P_i^4 + P_i^{1'} + P_i^{5'} + P_i^{6'}$$

$$= 0 + 0 + 0 + ne(x + \Delta d) - ne\left(x + \Delta d + \frac{x}{2}\right) - ne\left(x + \Delta d + \frac{x}{2}\right) \quad (2)$$

Note that compared with the initial state, the final state involves only the movement of site 1 (from 1 to 1') and two WCCs, 5 and 6 (from 5 and 6 to 5' and 6'). Sites 2 and 3, as well as WCC 4, remain stationary (Fig. 1b). Noting that $2x + \Delta d = a$ (see Fig. 1b), we obtain

$$P_f = -ne(2x + \Delta d) = -nea \quad (3)$$

Therefore, the dipole moment difference between the final and initial states is

$$\Delta P = P_f - P_i = -nea \quad (4)$$

The equation (4) confirms the presence of an integer-quantized dipole moment shift along the $a$-axis, despite the system's non-polar $D_{3h}$ symmetry. We refer to this nontrivial case as type-II IQFE, in contrast to the trivial type-I IQFE. In type-II IQFE, the initial and final structures are distinct and connected by a symmetry operation that does not belong to the symmetry group of either structure. Specifically, in the lattice model shown in Fig. 1b, the two states are related by mirror symmetry $M_x$, which is not a symmetry of the system itself. Notably, when considered separately, the displacements of ions or electronic charge centers are neither quantized nor fractional modulo the lattice constant $a$. However, their combined contribution results in an integer-quantized dipole moment. This behavior is fundamentally distinct from previously reported cases of fractional quantum ferroelectricity[28]. A particularly intriguing aspect of this kind of type-II IQFE is that the ionic displacement $\Delta d$ is a continuous parameter, and even a small $\Delta d$ can induce a quantized dipole moment due to a substantial shift in the electronic charge center. This is in stark contrast to conventional displacement-type ferroelectricity[37]—where the polarization is typically proportional to the magnitude of ionic displacements. In the following, unless otherwise specified, IQFE refers to type-II IQFE.

**IQFE in 3D chiral Te**

We propose that the aforementioned toy lattice model can be realized in the 3D Te system. Crystalline 3D Te belongs to the $D_3$ point group and exhibits structural chirality. This is because the $D_3$ point group consists only of proper rotation operations and does not contain any improper operations (e.g., inversion, mirror reflection, or rotoinversion), which are required to render a structure achiral. 3D Te exhibits two enantiomers, characterized by one-dimensional helical chains that spiral clockwise (CW) or anticlockwise (ACW) along the $c$-axis (see Figs. 2a and 2b). These chains interact with one another via van der Waals forces. Within each chain, each Te atom forms two covalent bonds with neighboring Te atoms (as shown in the side view in Fig. 2a), leaving two lone pairs and satisfying the octet rule[38]. To examine the electronic contribution to polarization, we calculate the WCCs for all valence bands of 3D Te using Wannier tight-binding models constructed from first-principles calculations (see Section I of Supplementary Material (SM)), implemented via the *Wannier90* package[39,40]. The WCCs are indicated by black spheres of Figs. 2a and 2b. Each Te atom contributes six valence electrons, leading to a total of eighteen valence electrons per primitive cell. Consequently, there are nine WCCs in the system, as each WCC hosts two electrons (spin–orbit coupling (SOC) is not included in WCCs calculation). By comparing the CW and ACW configurations (Figs. 2a and 2b), it is evident that the positions of the WCCs shift between the two states, indicating the presence of a net electronic polarization. The polarization difference contributed by the electrons is calculated to be -216.770 $\mu C/cm^2$, based on the displacement of the WCCs. The ionic contribution is computed to be 79.106 $\mu C/cm^2$, resulting in a total polarization of -

137.664 $\mu C/cm^2$. Notably, the polarization quantum for bulk Te is –68.848 $\mu C/cm^2$, and the total polarization is approximately twice this magnitude, with only a negligible deviation attributable to numerical error. This result is in excellent agreement with our theoretical prediction from Equation (4), where the integer $n = 2$ corresponds to the fact that each WCC contains two electrons (Thus the integer polarization value is 137.696 $\mu C/cm^2$). The ferroelectric energy barrier for bulk chiral Te is estimated to 15 meV per Te atom (Section II of SM). Using the MLP, we performed MD simulations under applied electric fields to examine the polarization response of bulk Te (Section III of the SM). The resulting hysteresis loop further confirms the ferroelectric nature of bulk Te, in agreement with our toy model analysis. In the following, we demonstrate how this ferroelectric behavior can be inherited by thin-film Te.

**Nearly IQFE in thin-film Te**

To explore the ferroelectric relationship between thin-film and bulk Te, we first examine the structural differences between them. The crystal structure of 2D Te is shown in Fig. 3a, where a six-layer configuration (supercell) is presented as a representative example. Compared to bulk Te, the structure of 2D Te remains largely unchanged due to the weak interlayer van der Waals interactions. It is evident that 2D Te can be obtained by cleaving bulk Te without disrupting the one-dimensional helical chains. This observation suggests that 2D Te should inherit the IQFE intrinsic to the bulk phase. However, it is worth noting that the point group of 2D Te is $C_2$, which is polar and thus permits conventional polarization along the $C_2$-axis. In this study, we define the $C_2$-axis as the $x$-direction (see Fig. 3a). Therefore, the total polarization in 2D Te should consist of two contributions: (i) the integer quantum polarization $P_Q$, inherited from the bulk Te, and (ii) the surface polarization $P_S$, permitted by the polar $C_2$ symmetry. This relationship can be expressed as:

$$P_T^{2D} = P_Q + P_S \quad (5)$$

Given that the quantum polarization is typically large, and the surface contribution is small, it is reasonable to expect that the total polarization $P_T^{2D}$ remains close to an integer quantum value. To verify this hypothesis, we performed calculations of the polarization in 2D Te with varying thicknesses (1, 2, 3, and 4 layers) using the WCCs method[41]. The results are presented in Fig. 3e. It can be observed that the total polarization $P_T^{2D}$ approaches the integer quantum polarization $P_Q$ as the layer thickness increases. For 2D Te, the polarization quantum is calculated to be -89.601 $\mu C/cm^2$, and $P_Q$, defined as twice the polarization quantum, is thus -179.201 $\mu C/cm^2$. As the thickness increases, the contribution from the surface polarization $P_s$ (in magnitude) decreases accordingly. Based on this trend, we conclude that with increasing thickness, the total polarization $P_T^{2D}$ asymptotically approaches the integer quantum polarization $P_Q$ (This is why we refer to it as nearly IQFE in thin-film Te), while the surface polarization $P_s$ tends toward zero. This behavior is consistent with the bulk limit, which can be regarded as the case of infinite layer thickness, where the surface polarization vanishes, and the total polarization equals $P_Q$. It is important to note that structural optimization was not performed prior to the polarization calculations using the WCCs method; hence, no surface atomic reconstruction effects are included. This observation suggests that the net surface polarization in 2D Te arises solely from electronic contributions, rather than ionic displacements. In addition, we repeated the polarization calculations using the WCC method for the optimized layered structures. As shown in Section IV of the SM, the results remain nearly unchanged, indicating that structural optimization has a negligible effect. This is reasonable, given the weak van der Waals interactions between the individual one-dimensional chains.

We have now demonstrated that the total polarization of 2D Te indeed consists of two components: the quantum polarization $P_Q$ and the surface polarization $P_s$. From a symmetry perspective, the emergence of the surface polarization in 2D Te originates from the breaking of the bulk Te's threefold rotational symmetry ($C_3$) during the exfoliation process. From a physical standpoint, this surface component primarily arises from surface charge redistribution, which results from the different local environments experienced by surface atoms compared to inner atoms—the latter being similar to those in the bulk structure. This interpretation is supported by our findings that $P_s$ increases as the layer thickness decreases (see Fig. 3e). Therefore, we conclude that in 2D Te, the integer quantum polarization $P_Q$ is inherited from bulk Te, while the $P_s$ arises due to surface charge redistribution caused by the breaking of the $C_3$ symmetry.

Using the MLP, we performed MD simulations under applied electric fields for four- to six-layer Te systems to investigate the polarization response (Section III of SM). For the six-layer Te simulation, the polarization value is very close to an integer quantum polarization (Section III of SM), confirming the nearly IQFE in thin-film Te. By carefully examining the switching process, we find that the surface atoms switch first, followed by the inner atoms, as illustrated by the snapshots in Fig. 3a–d. To understand this behavior, we perform two comparative calculations for six-layer 2D Te. In the first case, layers 2–6 are fixed while only the first layer is allowed to switch. In the second case, layers 1–2 and 4–6 are fixed while only the third layer is allowed to switch. We then compute the energy barriers between the initial and final states for both cases. The results are shown in Fig. 4a and Fig. 4b, respectively. It is evident that the surface layer exhibits a lower switching energy barrier compared to the inner layer, which explains why the surface layers switch earlier than the inner layers. The higher barrier observed in the inner layers can be attributed to the presence of more domains compared to the surface layers (see Fig. 4a and Fig. 4b). A greater number of domains implies that the system must overcome a larger energy barrier when the polarizations reverse direction.

Up to now, to our best knowledge, the switching of ferroelectric polarization of bulk Te has not been observed by any experiments. One possible reason is related to the switching mechanism. As previously demonstrated, the surface layers switch before the inner layers due to a lower energy barrier. In the case of bulk Te, the surface-to-volume ratio is significantly smaller, making the contribution of surface switching negligible. This explains why polarization switching is observed in thin films but not in bulk. Another contributing factor may be the difference in electronic properties between bulk and 2D (or thin-film) Te. Bulk Te possesses a smaller band gap (0.43eV from HSE06+SOC calculations) compared to 2D Te[33] (Section V of SM), which suggests that bulk Te experiences higher leakage currents. This elevated leakage current can inhibit successful polarization switching in bulk Te. On the other hand, the bulk selenium (Se), which adopts the same structure as bulk Te and has been experimentally synthesized, exhibits a larger band gap (about 1.72 eV, Section V of SM), potentially indicating lower leakage currents. In addition, the energy barrier between the two polarization states in bulk Se is estimated to be 84 meV per atom, which is not large compared to the $BaTiO_3$ (about 60 meV)[42], suggesting that polarization switching can be driven by an electric field.

**Natural optical activity in IQFE chiral Te**

Natural optical activity (NOA)—the rotation of the polarization plane of light as it propagates through a medium[43,44], serves as a powerful optical probe for detecting symmetry-breaking phenomena. For IQFE Te ($D_3$ point group), the gyrotropic tensor $g_{\alpha\beta}(\omega)$ has the following form[45]:

$$g_{\alpha\beta}(\omega) = \begin{bmatrix} g_{11}(\omega) & 0 & 0 \\ 0 & g_{11}(\omega) & 0 \\ 0 & 0 & g_{33}(\omega) \end{bmatrix} \quad (6)$$

Here, $\omega$ denotes the angular frequency of the incident light. It is note that the optical rotatory power $\rho$ depends solely on the $g_{33}$[43], i.e.,

$$\rho(\omega) = \frac{\omega^2}{2c^2} g_{33}(\omega) \quad (7)$$

Using density-functional perturbation theory in ABINIT [46-48], we calculate $g_{33}$ for opposite chiral configurations, yielding +2.05 bohr (left-) and –2.05 bohr (right-handed Te), notably larger than that of α-quartz (0.125 bohr)[46]. This result indicates that left- and right-handed Te exhibit opposite optical rotation when light propagates along the optical axis (the $z$-direction). Importantly, Chiral Te displays ferroelectric-like behavior, with an electric-field-controllable polarization and structural chirality. This enables deterministic switching between left- and right-handed crystal domains. The switchable structural chirality results in a sign reversal of the optical rotation, offering a robust mechanism for electric writing of chiral states and optical reading via rotation measurements. Note that previous studies have demonstrated that an electric field can control the chirality of polar vortices[49], which arises from the helicity of dipole moments in polar materials with polar space groups, even when the crystal structure itself is achiral. In contrast, we propose electric-field control of structural chirality in a material with a non-polar space group—a fundamentally different mechanism. Such electrically tunable optical activity in Te opens new avenues for chiral photonics and nonvolatile memory devices that integrate electric control with optical functionality.

**Conclusions**

We have theoretically uncovered a new class of ferroelectric order—type-II IQFE. Through a toy model and first-principles WCCs analysis, we demonstrated that an integer quantum polarization difference emerges between the two enantiomeric forms of bulk chiral Te. The total polarization, locked to integer multiples of the quantum of polarization, distinguishes this mechanism from both conventional and fractional quantum ferroelectrics. In thin-film Te, the polarization remains nearly quantized, with a surface-induced conventional component diminishing with thickness. MD simulations reveal a layer-by-layer switching mechanism, initiated by surface atoms with lower energy barriers. Furthermore, the two chiral states exhibit opposite optical rotation, and their electric-field-controlled interconversion enables deterministic modulation of NOA. These findings establish Te as a rare system where electric, structural, and optical chiralities are intrinsically coupled and switchable, offering new opportunities for quantum ferroelectricity and electrically programmable chiral photonics.

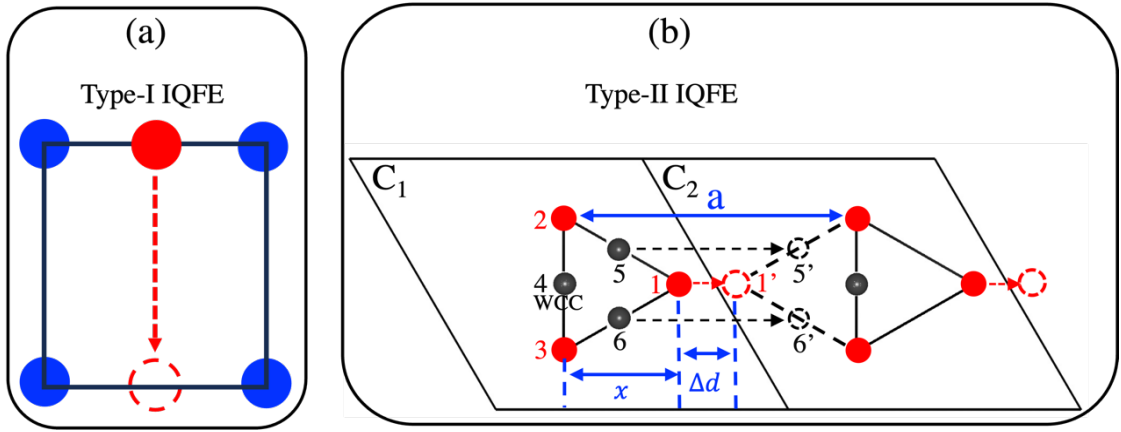

Fig. 1 (a) The schematic diagram illustrates the trivial case of IQFE. The red and blue dots represent the anions and cations, respectively. (b) Toy lattice model with $D_3$ symmetry. Site 1 in cell $C_1$ shifts by $\Delta d$ to site 1' in cell $C_2$, resulting in a jump of the WCCs (from 5 and 6 to 5' and 6').

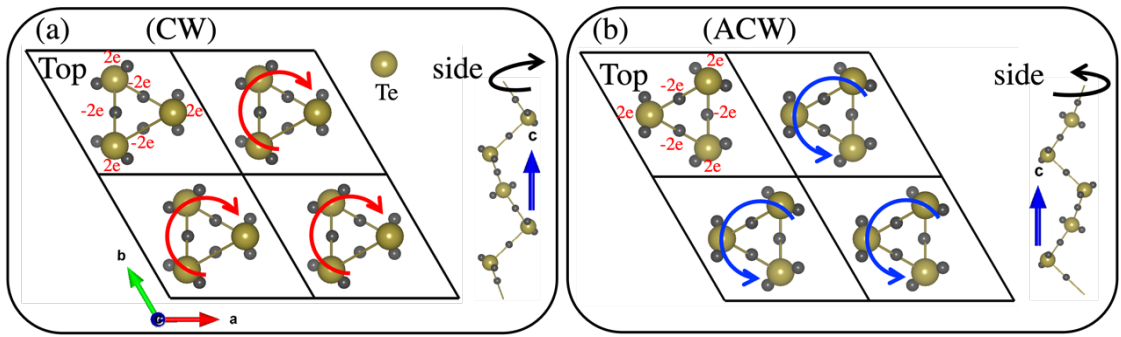

Fig. 2 Top and side views of bulk Te in CW (a) and ACW configuration (b). Black spheres denote WCCs.

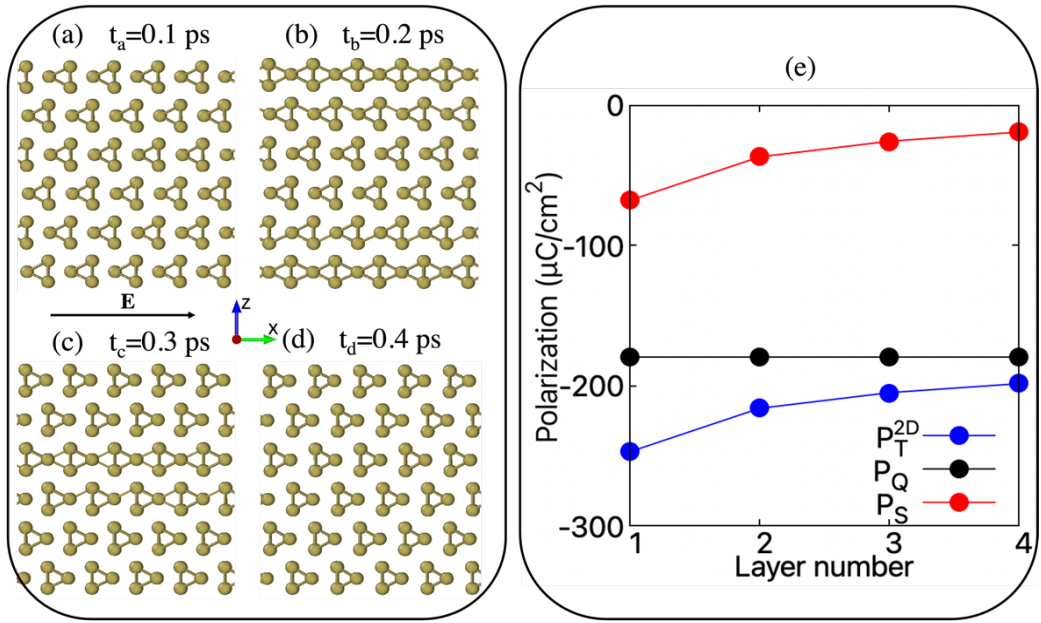

Fig. 3 (a)–(d) Snapshots of the switching process at successive time steps. (e) Variation of $P_T^{2D}$ and $P_S$ with the number of layers in 2D Te.

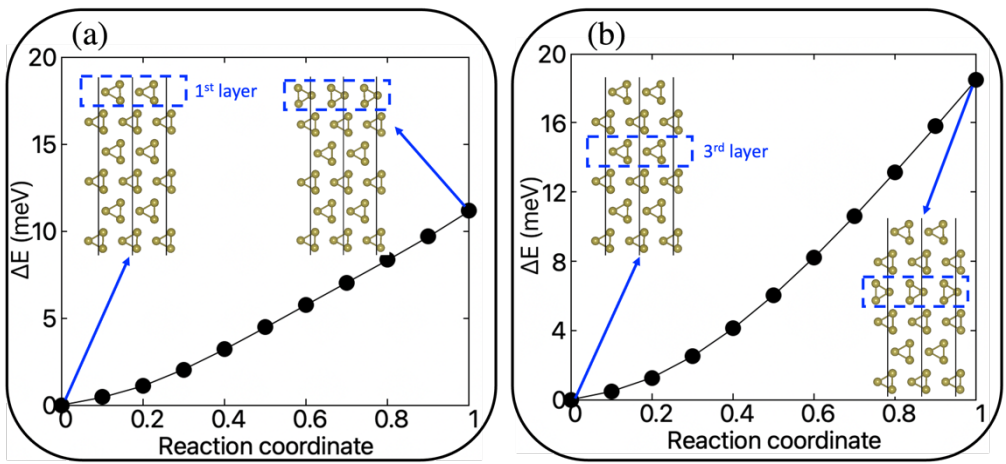

Fig. 4 Energy barrier for the 1st layer (a) and the 3rd layer (b). The barriers are 11.1 meV and 18.5 meV per atom for the 1st and 3rd layers, respectively.

**Acknowledgments**

Work at Fudan is supported by NSFC (grants No. 12188101), the National Key R&D Program of China (No. 2022YFA1402901), Shanghai Science and Technology Program (No. 23JC1400900), the Guangdong Major Project of the Basic and Applied Basic Research (Future functional materials under extreme conditions--2021B0301030005), Shanghai Pilot Program for Basic Research—FuDan University 21TQ1400100 (23TQ017), the robotic AI-Scientist platform of Chinese Academy of Science, and New Cornerstone Science Foundation. This research at UARK is supported by the Arkansas High Performance Computing Center (AHPCC) which is funded through multiple National Science Foundation grants and the Arkansas Economic Development Commission. W.L. and L.B. received support from a Vannevar Bush Faculty Fellowship (VBFF), Grant No. N00014-20-1-2834 from the Department of Defense. J. J. also acknowledges the support from China National Postdoctoral Program for Innovative Talents (BX20230408).